\documentclass[%
 reprint,
 superscriptaddress,
 amsmath,amssymb,
 aps,
]{revtex4-2}

\usepackage{graphicx}
\usepackage{dcolumn}
\usepackage{bm}

\usepackage{color}
\usepackage{booktabs}

\begin{document}

\preprint{APS/123-QED}

\title{Reproduction of NGC1052-DF4 by self-interacting dark matter:\\ dark matter deficiency and tidal features}

\author{Zhao-Chen Zhang$^*$}
\affiliation{Key Laboratory of Particle Astrophysics, Institute of High Energy Physics,
Chinese Academy of Sciences, Beijing 100049, China}

\affiliation{School of Physical Sciences, University of Chinese Academy of Sciences, Beijing, China}

\author{Xiao-Jun Bi$^\dag$}
\affiliation{Key Laboratory of Particle Astrophysics, Institute of High Energy Physics,
Chinese Academy of Sciences, Beijing 100049, China}

\affiliation{School of Physical Sciences, University of Chinese Academy of Sciences, Beijing, China}

\author{Peng-Fei Yin$^\P$}
\affiliation{Key Laboratory of Particle Astrophysics, Institute of High Energy Physics,
Chinese Academy of Sciences, Beijing 100049, China}

\date{\today}

\begin{abstract}
Observations of the velocity dispersion indicate a severe dark matter (DM) deficit in the ultra-diffuse galaxy, NGC1052-DF4 (DF4).
The ultra-deep images obtained with the Gemini telescope, which has the deepest imaging data till now, confirm the presence of tidal tails in DF4, suggesting its tidal formation.
To enhance tidal effects, we consider the self-interaction among DM particles.
Using an N-body simulation in the scenario of self-interacting dark matter (SIDM), we reproduce a DM-deficient galaxy that is consistent with all observational data of DF4.
Specifically, our simulation result yields an extremely low DM-to-star mass ratio and a radial surface brightness profile very similar to that from deep images, showing accurate tidal features.
By performing simulations with similar tidal effects and various cross-sections of SIDM, we show a significant impact of SIDM on the DM-to-star mass ratio in the central region of the galaxy.
Our work confirms the tidal formation of DF4 in theory.
\end{abstract}

\maketitle


\section{\label{sec:introduction} Introduction}

Ultra-diffuse galaxies (UDGs) are a type of galaxies characterized by a diffuse spatial distribution of stars and low luminosity \cite{vanDokkum:2014cea}. 
Owing to the limited influence of baryonic matter, UDGs provide an optimal environment for exploring the intrinsic properties of dark matter (DM).
However, the observed low velocity dispersion of NGC1052-DF2 (DF2), which is identified as a UDG, implies a scarcity or absence of DM within this galaxy \cite{vanDokkum:2018vup}.
Subsequently, a second UDG, NGC1052-DF4 (DF4), exhibiting similar characteristics to DF2, is discovered within the same region \cite{van2019second}. The kinematics of DF4 indicate an even more severe deficiency of DM compared to DF2.
The discovery of these two UDGs raises the possibility of a population of DM-deficient galaxies and points towards a coherent explanation for their formation.

One plausible mechanism for this observed phenomenon is tidal stripping, a process which removes DM from the outer regions of the halo, while exerting less influence on the central baryonic matter.
Controlled N-body simulations have demonstrated that a UDG resembling DF2 can be generated through tidal stripping by its host halo \cite{Ogiya:2018jww, Yang:2020iya, Ogiya:2021wyz}.
However, the observation of tidal features presents distinct results for DF2 and DF4.
Deep imaging conducted with the Isaac Newton Telescope (INT) reveals no evidence of tidal distortion in DF2 \cite{Montes:2021njd}.
In contrast, Ref. \cite{Montes:2020zaz} and \cite{keim2022tidal} have discovered tidal tails in DF4, with the IAC80 telescopes and the Dragonfly Telephoto Array, respectively.
Recently, ultra-deep images of DF2 and DF4 captured by the Gemini telescopes, which are approximately 1 mag deeper than previously available data, have confirmed the absence of tidal features in DF2 and the presence of tidal tails in DF4 \cite{golini2024ultra}.

It is challenging to interpret the formation of DF4 with tidal effects in the framework of the standard cold dark matter (CDM) model.
On one hand, DF4 exhibits a more severe DM deficit compared to DF2.
Specifically, the ratio of DM to stellar mass enclosed within $R=7~\rm kpc$ has an upper limit of 0.45 for DF4 \cite{van2019second}, while this ratio has an upper limit of 1.65 at $R=2.7~\rm kpc$ in DF2 \cite{Danieli:2019zyi}.
On the other hand, the projected distance from DF4 to NGC1052, which is considered as the host galaxy of DF4 and DF2 in many studies \cite{van2019second, danieli2020tip, keim2022tidal}, is approximately twice that of DF2 \cite{vanDokkum:2018vup, van2019second}.
This suggests that DF4 may be following a less bound orbit, resulting in weaker tidal effects.
Therefore, in order to explain the formation of DF4, additional mechanisms are necessitated to enhance tidal effects.

Self-interactions among DM particles can potentially enhance tidal effects experienced by a satellite galaxy \cite{Yang:2020iya}. 
In the self-interacting dark matter (SIDM) model, the colder DM particles within the inner halo undergo thermalization due to self-interactions, leading to their outward motion \cite{Spergel:1999mh, Burkert:2000di, Dave:2000ar, Vogelsberger:2012ku, Rocha:2012jg}.
This process results in a shallower gravitational potential, making the halo more vulnerable to tidal forces.
Thus the tidal effects are enhanced by DM self-interaction.
The SIDM model also provides interpretations for various observational discrepancies with the $\Lambda$CDM paradigm, such as the diversity problem in rotation curves \cite{Kamada:2016euw, Ren:2018jpt, Kaplinghat:2019dhn} and the too-big-to-fail problem \cite{Zavala:2012us, Elbert:2014bma, Vogelsberger:2012ku}.

In this study, we utilize controlled N-body simulations in the framework of SIDM to explore whether the enhanced tidal stripping can effectively replicate a galaxy similar to DF4.
Our investigation focuses on two crucial aspects of DF4: the extremely low DM-to-star mass ratio, and the obvious tidal features in the surface brightness profile.
We expect that the extreme tidal evolution inferred from the observations of DF4 may provide valuable insights into the intrinsic properties of DM.

This paper is organized as follows.
In Section~\ref{sec:simulation}, we introduce the details of our simulation setup, encompassing the selection of initial conditions and the numerical implementation of DM self-interactions.
We outline our methodology and approach to modeling the system.
The simulation outcomes and pertinent discussions are presented in Section~\ref{sec:results}, where we analyze the results obtained from our simulations and interpret their implications in relation to the observed properties of DF4.
Finally, we provide a summary in Section~\ref{sec:summary}.

\section{\label{sec:simulation}Simulation}

We conduct a fiducial simulation in the framework of SIDM to reproduce a galaxy similar to DF4.
The host system is characterized as an accreting DM halo resembling that of NGC1052, and is modeled by an analytical time-varying potential background in our simulations to enhance computational efficiency. 
The DM particles and star particles constitute the N-body satellite system.
Given that the stellar population of DF4 has an age of $8.76^{+2.91}_{-1.51}~\rm Gyr$ \cite{buzzo2022stellar}, our fiducial simulation starts at $z_{\rm i}=1.5$, corresponding to a look-back time of $t=\rm -9.54~Gyr$, and proceeds to $z_{\rm f}=0$.

\subsection{\label{subsec:sidm}Self-interaction implementation and numerical parameters}

In this work, we consider a general scenario of velocity-dependent DM scattering, following the model in \cite{Kaplinghat:2015aga}.
Various DM halos, spanning different scales, exhibit a range of mean collision velocities for DM particles, resulting in diverse cross-sections.
The cross-section diminishes as the mean collision velocity increases. 
Nonetheless, the scatterings within a single halo still maintain a constant cross-section which is decided by the average collision velocity within this halo.
According to this model, we assume elastic and isotropic scatterings among the DM particles within the satellite halo, with a constant cross-section.
It is worth noting that the cross-section of DM scatterings within the host halo is reduced due to the velocity-dependent mechanism.
Therefore, the formation of a density core in the host halo and the evaporation of the satellite halo can be neglected (see also in Section~\ref{subsec:host}).
Cross-section values ranging from $0.3~\rm cm^{2}/g$ to $10.2~\rm cm^{2}/g$ have been employed to account for the central density cores observed in certain dwarf galaxies \cite{Kaplinghat:2015aga}, thus we consider this range as a reasonable interval. 
To enhance the tidal effects and achieve an extremely low ratio of DM to stellar mass, we utilize a SIDM cross-section of $\sigma/m=10~\rm cm^{2}/g$ in our fiducial simulation, closely approaching the upper limit of the reasonable interval.

We have made modifications to the widely used N-body simulation code \texttt{GADGET-2} \cite{Springel:2005mi,Springel:2000yr}, incorporating self-interactions among DM particles via a Monte-Carlo method, as outlined in \cite{Robertson:2016xjh}.
The density profile of a single halo produced by our adapted code is consistent with the results obtained from the semi-analytical model in \cite{Kaplinghat:2015aga}, thus verifying the accuracy of our SIDM implementation.
In our simulations, both DM and star particles possess a mass of $10^{4}~\rm M_{\odot}$ and a softening length of $40~\rm pc$.
Note that this resolution is sufficiently high to avoid non-physical effects on the scales of our investigation \cite{vandenBosch:2017ynq}.
The initial condition files for the N-body satellite system are generated by the public code \texttt{SpherIC} \cite{Garrison-Kimmel:2013yys}.

\subsection{\label{subsec:host}Host system}

Observations indicate that the total stellar mass of NGC1052 is $M_{\star}=10^{11}~\rm M_{\odot}$ \cite{forbes2017sluggs}.
According to the stellar-halo mass relation \cite{Moster:2012fv}, the current halo mass of NGC1052 is estimated to be $M_{200}=1.1\times10^{13}~\rm M_{\odot}$.
We model the host halo with the NFW profile \cite{Navarro:1996gj}.
The configuration of the halo is characterized by its viral mass $M_{\rm 200}$ and concentration parameter $c_{200}=r_{200}/r_{\rm s}$, where $r_{200}$ and $r_{s}$ are the virial and scale radius, respectively.
By incorporating the mass accretion history \cite{Correa:2015kia} and the redshift-dependent mass-concentration relation \cite{Ludlow:2016ifl}, the $(M_{200},c_{200})$ configuration of the host halo evolves from $(3.6\times10^{12}~\rm M_{\odot},4.8)$ at $z_{\rm i}=1.5$ to $(1.1\times10^{13}~\rm M_{\odot},6.8)$ at $z_{\rm f}=0$.

In our model of the host system, certain approximations are inevitable.
The analytical potential eliminates the dynamical friction and evaporation of the satellite halo.
Since the timescale for orbit decay caused by dynamical friction is significantly longer than the age of the universe \cite{Ogiya:2018jww, Yang:2020iya}, the dynamical friction can be safely neglected.
Scatterings between DM particles of the host halo and those of the satellite halo result in the evaporation of the satellite halo.
Considering the velocity-dependent scattering discussed earlier, the cross-section of this scatter is $\sigma/m=0.4~\rm cm^{2}/g$ when the satellite passes the pericenter at a velocity of $v_{\rm peri}=651~\rm km/s$.
This weak interaction is expected to induce minimal mass loss \cite{Silverman:2022bhs}, thus the absence of evaporation would have negligible influences on our simulations.

Due to the lack of the stellar evolution history, we omit the stellar component within the host system and instead conduct a conservative evaluation of the tidal effects.
Additionally, we neglect the density core in the host halo induced by DM self-interactions.
The mean collision velocity of DM particles within the host halo is $v=507~\rm km/s$, corresponding to a potential cross-section $\sigma/m=0.6~\rm cm^{2}/g$ \cite{Kaplinghat:2015aga}.
DM scatterings with this cross-section can generate a density core with a radius of $R_{\rm c}=16.9~\rm kpc$ until $z_{\rm f}=0$ \cite{Jiang:2022aqw}.
At the pericenter distance of $R_{\rm peri}=20.1~\rm kpc$, the presence of the core results in a reduction in the enclosed DM mass by approximately $4.1\times10^{10}~\rm M_{\odot}$, compared to the NFW profile.
Notably, this reduction in mass is lower than the stellar mass of NGC1052, indicating that the enclosed mass at the pericenter in our model is lower than that in the actual host system.
Therefore, our model of the host system can be regarded as a conservative estimate of the actual tidal effects.

\subsection{\label{sub:satellite}Satellite system}
In our simulations, we set the initial halo mass of the satellite system to be $M_{200}=8.0\times10^{10}~\rm M_{\odot}$.
We also adopt a stellar mass of $M_{\star}=5.2\times10^{8}~\rm M_{\odot}$, which is $1.6\sigma$ higher than the median value from the stellar-halo mass relation \cite{Behroozi:2019kql}.
After multiple attempts, we have found that it is difficult for a typical satellite system to evolve into a DF4-like galaxy with a reasonable SIDM cross-section.
Therefore, to obtain an extremely DM deficit galaxy, the initial satellite system should consist of a diffuse DM halo and a compact star population.
We model the initial satellite halo using the NFW profile with a concentration parameter of $c_{200}=3.3$, which is $2\sigma$ lower than the median value derived from the redshift-dependent mass-concentration relation \cite{Dutton:2014xda}.
The initial distribution of stars is represented by the Plummer profile \cite{Plummer:1911zza}, with an effective radius of $R_{\rm e}=1.0~\rm kpc$, which is $2.1\sigma$ lower than the median value derived from the mass-size relation of galaxies \cite{carleton2019formation}.

\subsection{\label{orbit} Orbit}
The orbit of the satellite system is characterized by two dimensionless parameters: the energy parameter $x_{\rm c}$ and the circularity parameter $\eta$.
The energy parameter is defined as $x_{\rm c}\equiv r_{\rm c}(E)/r_{200}$, where $r_{\rm c}(E)$ is the radius of a circular orbit with the same energy $E$, and $r_{200}$ is the virial radius of the host halo.
The circularity parameter is defined as $\eta\equiv L/L_{\rm c}(E)$, where $L$ and $L_{c}$ denote the angular momentum of the orbit and that of a circular orbit with the same energy $E$, respectively.

The projected distance from the center of NGC1052 to DF4 is estimated to be $165~\rm kpc$ \cite{van2019second}, which serves as a lower limit for the 3D distance.
We adopt an orbit with $x_{\rm c}=1.0$ and $\eta=0.25$. 
The satellite halo is initially positioned at the apocenter with a distance of $R_{\rm apo}=307~\rm kpc$ and a velocity of $v_{\rm apo}=42.6~\rm km/s$.
This initial apocenter distance is sufficiently large to ensure that the reproduced DF4-like galaxy can locate at a distance beyond 165 kpc, even considering the shrinkage of orbit caused by the accretion of the host halo.
The pericenter distance is $R_{\rm peri}=20.1~\rm kpc$, corresponding to the 7.5 percentile of the distribution provided by \cite{vandenBosch:2017ynq}, indicating that our chosen orbit parameters are not overly extreme.

\section{\label{sec:results}Results and discussion}
Our fiducial simulation successfully reproduces a galaxy resembling DF4 at a look-back time of $t=-0.34~\rm Gyr$.
In addition to the fiducial simulation, we conduct two supplementary simulations with different combinations of SIDM cross-section and orbit to demonstrate the important role of DM self-interaction in reducing the DM-to-star mass ratio in the central region.
Furthermore, a separate simulation, in which the initial star distribution is modeled using the Hernquist profile, is performed to illustrate the influence of the initial distribution on the ultimate surface brightness profile.
Note that we assume the distance of DF4 is $D=20.0\pm1.6 \rm~Mpc$ in all our analyses \cite{danieli2020tip}.

\begin{figure}[tp!]
\includegraphics[width=\linewidth]{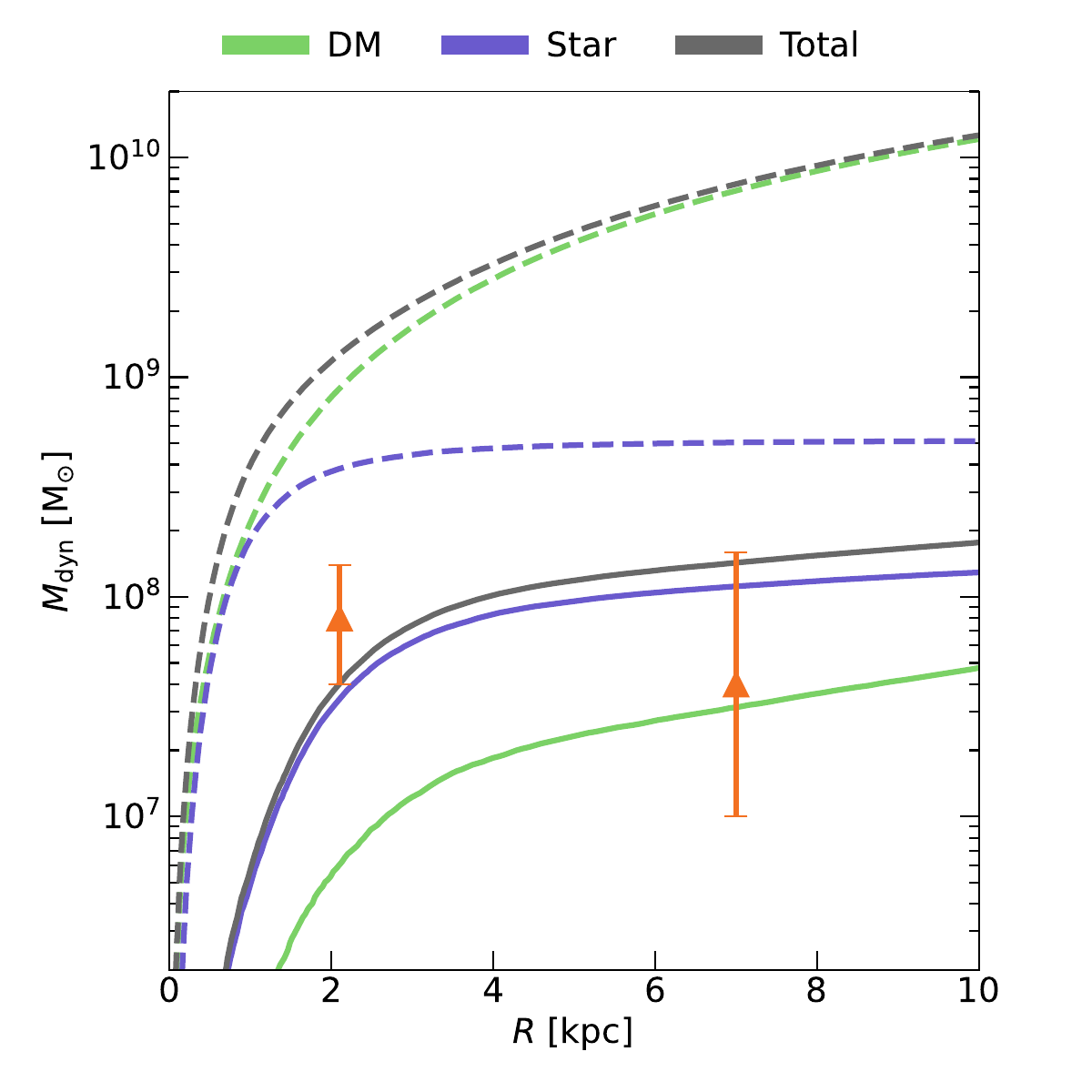}
\caption{\label{mass_profile}
The dynamical mass profile of the satellite system.
The solid lines represent profiles at $t=-0.34~\rm Gyr$, when the results of the fiducial simulation best match the observational data.
The dashed lines represent the initial profiles.
The orange data points with error bars at $R=2.1~\rm kpc$ and $7~\rm kpc$ denote total dynamical mass of $M_{\rm dyn}=0.8^{+0.6}_{-0.4}\times10^{8}~\rm M_{\odot}$ and $0.4^{+1.2}_{-0.3}\times10^{8}~\rm M_{\odot}$ enclosed within corresponding radii, obtained by the velocity dispersion measurement of stars \cite{Shen:2023jwk} and GCs \cite{van2019second} in DF4, respectively.}
\end{figure}

\subsection{\label{subsec:mass_profile}Mass profile}

Fig. \ref{mass_profile} illustrates the dynamical mass profile at $t=-0.34~\rm Gyr$, when the results of our fiducial simulation are consistent with all observational data of DF4.
The initial profiles are also plotted as a comparison.
Undergoing intense tidal stripping enhanced by DM self-interactions, the DM halo of the satellite galaxy loses approximately $99.95\%$ of its bound mass.
While the mass loss of stars is also significant, it is considerably smaller in percentage compared to that of the DM. 
Specifically, the final bound mass of stars is $25\%$ of its initial value.
This discrepancy can be attributed to the concentration of stars in the central region of the satellite system, resulting in less impact from tidal forces overall. 
Additionally, DM self-interactions play a crucial role.
The kinetic energy of the hot outer halo is transmitted inwards through self-interactions, causing the inner DM particles to move outwards.
This process weakens the resistance of the halo to tidal forces and accelerates the mass loss directly.

Under the influence of tidal stripping and DM self-interaction, the DM halo eventually evolves into a highly diffuse distribution, characterized by a remarkably gradual slope in its mass profile.
Using the velocity dispersion of the globular clusters (GCs) in DF4, Ref. \cite{van2019second} reported a total dynamical mass of $M_{\rm dyn}=0.4^{+1.2}_{-0.3}\times10^{8}~\rm M_{\odot}$ within $R=7~\rm kpc$.
Based on this, Ref. \cite{Shen:2023jwk} measured the velocity dispersion of the stars in DF4 and suggested an enclosed dynamical mass of $M_{\rm dyn}=0.8^{+0.6}_{-0.4}\times10^{8}~\rm M_{\odot}$ within $R=2.1~\rm kpc$.
Notably, the mass profile derived from our fiducial simulation is in accordance with these observational data.

\begin{figure}[tp!]
\includegraphics[width=\linewidth]{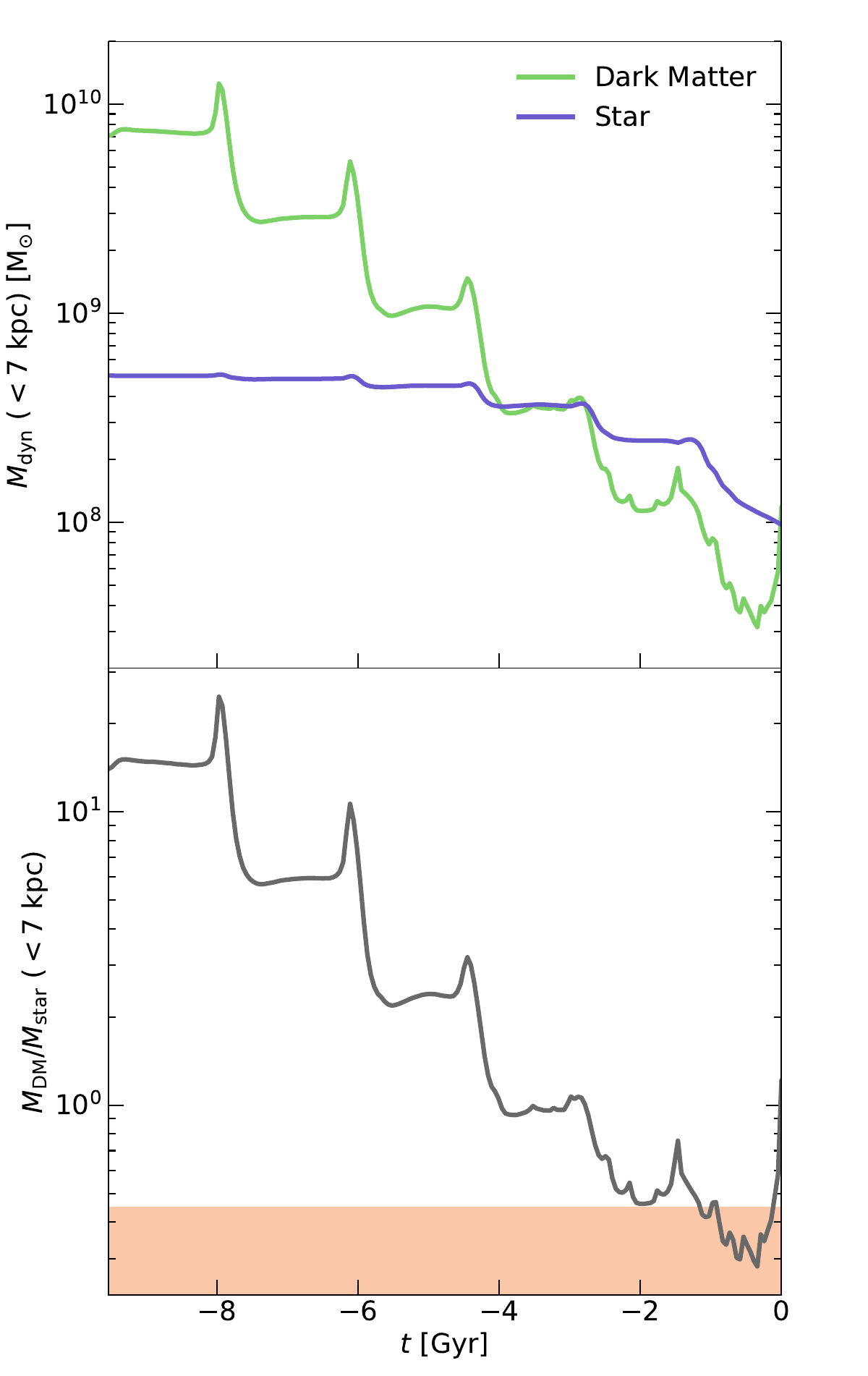}
\caption{\label{ratio}
\textit{Top}: The evolution of dynamical mass enclosed within $R=7~\rm kpc$.
The purple and green lines indicate the DM and star mass, respectively.
\textit{Bottom}: The ratio of DM to star mass enclosed within $R=7~\rm kpc$.
The light orange region denotes the interval of the mass ratio inferred from observational results (see the text for details).
Note that this interval is $0\sim0.45$, and the shaded region does not fully represent the whole interval due to the logarithmic scale.}
\end{figure}

\begin{figure}[tp!]
\includegraphics[width=\linewidth]{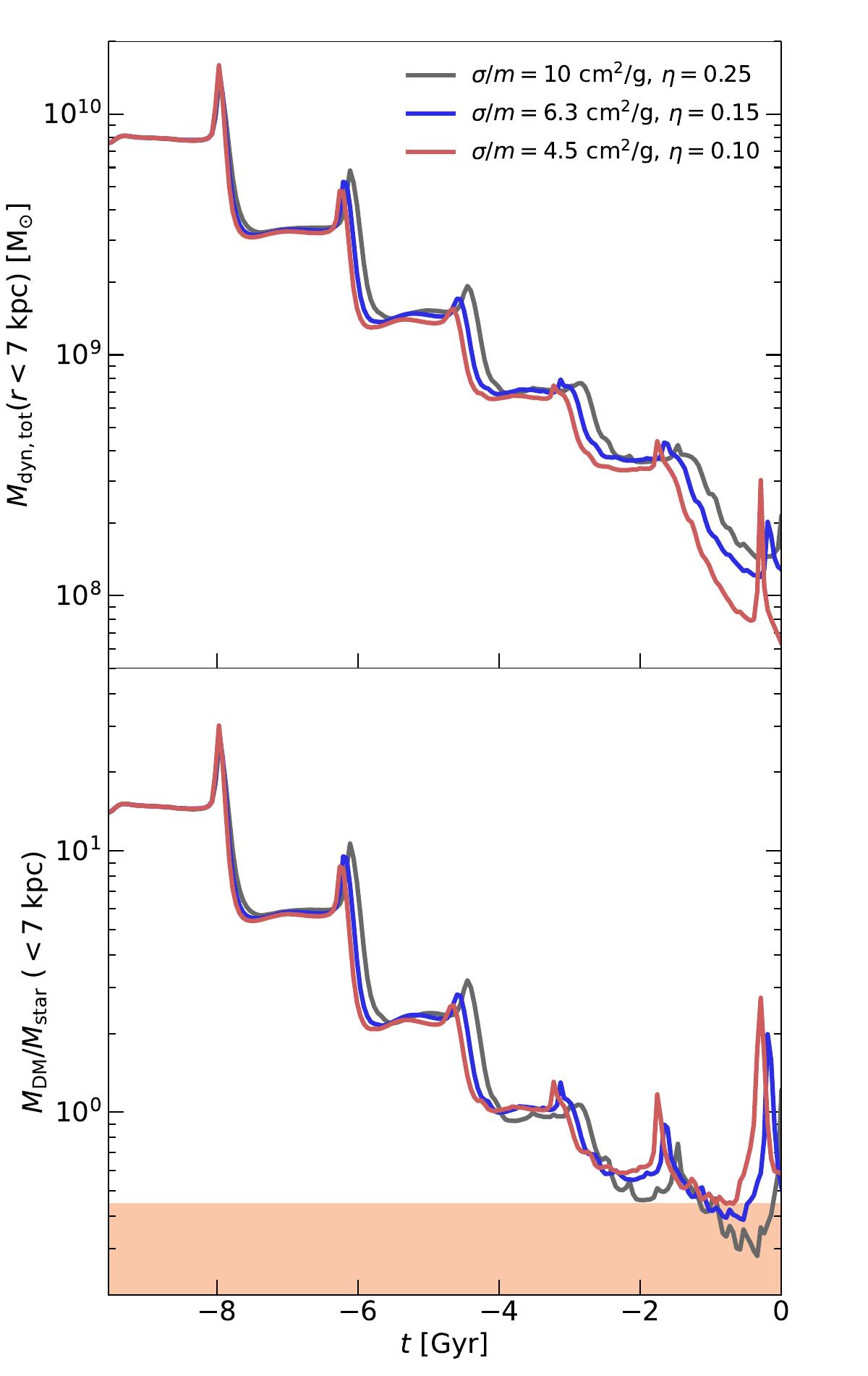}
\caption{\label{ratio_sidm}
\textit{Top}: The evolution of the total dynamical mass enclosed within $R=7~\rm kpc$.
\textit{Bottom}: The ratio of DM to star mass enclosed within $R=7~\rm kpc$.
Similar in Fig. \ref{ratio}, the light orange region denotes the allowable interval based on observations.
In this figure, lines of different colors represent results from simulations with different combinations of SIDM cross-section $\sigma/m$ and orbital circularity parameter $\eta$.
While the lines of the same color represent results from the same simulation.
Outcomes from our fiducial simulation are denoted by gray lines.}
\end{figure}

\subsection{\label{subsec:ratio}Mass ratio in the central region}

Fig. \ref{ratio} demonstrates the dynamical mass of DM and stars enclosed within $R=7~\rm kpc$, and the ratio between them.
Hereafter, the term ‘mass ratio' denotes the dynamical mass ratio of DM to stars enclosed within $R=7~\rm kpc$.
Both components exhibit a decreasing trend in mass.
The enclosed DM mass declines at a notably swifter pace, leading to a rapid decrease in the mass ratio.
Ref.~\cite{cohen2018dragonfly} reported that the stars in DF4 possess a total mass of $M_{\star}=(1.5\pm0.4)\times10^{8}~\rm M_{\odot}$ and follows a S$\rm \acute{e}$rsic profile with an index of $n=0.79$.
The mass of stars enclosed within $R=7~\rm kpc$ constitutes over $99\%$ of the total stellar mass in this profile.
In combination with the total dynamical mass $M_{\rm dyn}=0.4^{+1.2}_{-0.3}\times10^{8}~\rm M_{\odot}$ \cite{van2019second}, the mass ratio is estimated to fall within the range of $0\sim0.45$.
From Fig. \ref{ratio}, it can be seen that our simulation results satisfy the constraints of the mass ratio.

The initial configuration of the satellite system can account for the decrease in the mass ratio.
The initial DM halo is diffuse with a low concentration parameter, while the distribution of stars is compact with a low effective radius.
The higher density and deeper gravitational potential of the stars in the central region enable them to resist tidal forces more effectively, resulting in a much slower rate of mass loss compared to DM.
Therefore, under the premise of such an initial configuration, tidal forces continuously decrease the mass ratio, with stronger tidal effects leading to a more pronounced reduction in the mass ratio.

It is worth noting that DM self-interaction also significantly contributes to the decrease in the mass ratio.
The thermalization of DM particles in the central region induces their outward motion, directly leading to a reduction in the total DM mass.
The mass ratio would decrease through this mechanism even without tidal effects.
To demonstrate the impact of DM self-interaction, we conduct two additional simulations.
In these simulations, we vary the SIDM cross-section $\sigma/m$ and the orbital circularity parameter $\eta$ to achieve a similar total dynamical mass, while keeping all other conditions consistent with the fiducial simulation.
The total dynamical mass and the mass ratio produced by these three simulations are shown in Fig. \ref{ratio_sidm}.
It can be seen that simulations with more radial orbits exhibit lower total dynamical mass, indicating stronger tidal effects.
Evolving from identical initial conditions, these simulations with stronger tidal effects are expected to produce lower mass ratios.
However, due to stronger DM self-interactions, simulations with weaker tidal effects more effectively reduce the mass ratio, yielding lower mass ratios in the later stages of evolution.
Thus the significance of DM self-interaction in this process is well illustrated.

In summary, the exceptionally low mass ratio is a combined result of the initial configuration influenced by the tidal field and the DM self-interaction.

\begin{figure}[tp!]
\includegraphics[width=\linewidth]{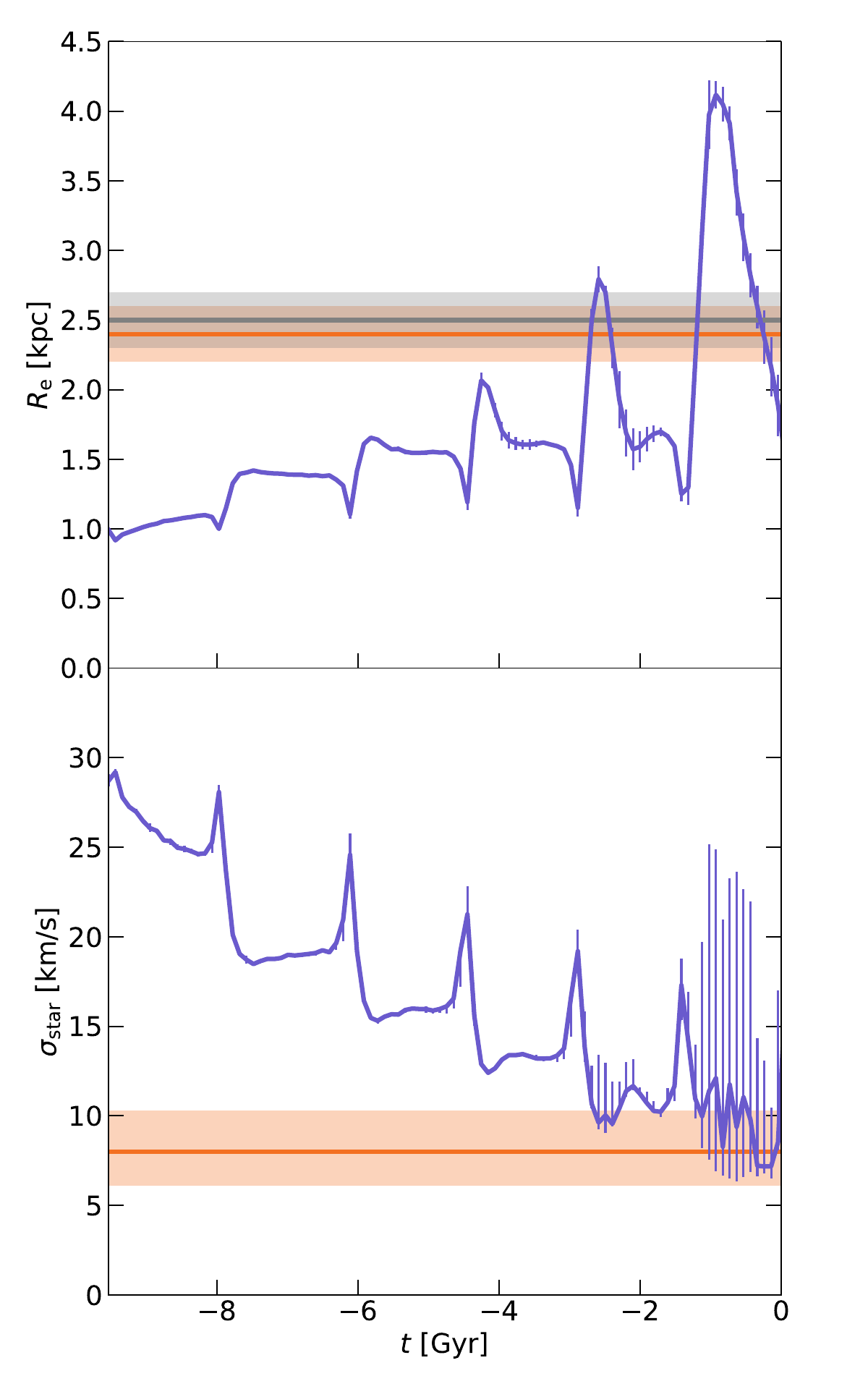}
\caption{\label{star_distribution}
\textit{Top}: The evolution of the effective radius.
The solid orange and gray lines denote effective radii of $R_{\rm e}=2.4~\rm kpc$ and $2.5~\rm kpc$, obtained by fitting isophotes with fixed and free ellipses, respectively \cite{golini2024ultra}.
The light orange and gray regions mark the corresponding error ranges introduced by the distance $D=20.0\pm1.6~\rm Mpc$ of DF4 \cite{danieli2020tip}.
\textit{Bottom}: The line-of-sight velocity dispersion among the stars enclosed within the effective radius as a function of time.
The solid orange line and the light orange region represent an observed velocity dispersion of $\sigma_{\rm star}=8.0^{+2.3}_{-1.9}~\rm km/s$ \cite{Shen:2023jwk}.
In both panels, the error bars mark the 15 to 85 percentile interval for the values obtained by projecting the stars using 100 random orientations.}
\end{figure}

\subsection{\label{subsec:star}Star distribution}

The evolution of the effective radius and the velocity dispersion of the stars is shown in Fig. \ref{star_distribution}.
The observed effective radius depends on the surface brightness limit of imaging, with deeper imaging providing a more accurate and larger effective radius measurement.
The imaging of Gemini, which provides the deepest images of DF4 till now, reports effective radii of $R_{\rm e}=24.5''$ and $25.8''$, determined by fitting isophotes using fixed ellipses and free ellipses, respectively \cite{golini2024ultra}.
Considering the distance of DF4, translating these angular measurements to physical scales yields length radii of $R_{\rm e}=2.4\pm0.2~\rm kpc$ and $2.5\pm0.2~\rm kpc$, respectively.
Ref. \cite{Shen:2023jwk} suggested a velocity dispersion of $\sigma_{\rm star}=8.0^{+2.3}_{-1.9}~\rm km/s$ for the star population in DF4.
In this study, we calculate the effective radius and the velocity dispersion assuming circular isophotes.
Since the ellipticity of DF4 is below 0.2 within $R=60''$ ($5.8~\rm kpc$) \cite{Montes:2020zaz,keim2022tidal,golini2024ultra}, our simplification is valid.
The star population in our fiducial simulation closely aligns with observational data in terms of velocity dispersion and effective radius, reinforcing the consistency of our simulation with the observed properties of DF4.

The evolution of the stars in the simulation displays prominent tidal features.
The tidal heating, induced by the radial velocity of the satellite system, leads to the contraction of the star population and an increase in temperature.
The sharp decrease in effective radius and the concurrent sudden rise in velocity dispersion serve as indicators of tidal heating.
Subsequently, due to the negative heat capacity of a gravitational system, the star population expands and cools to a lower temperature compared to the pre-heated phase.
Together with tidal stripping, the combined tidal effects contribute to an overall trend in which the star population exhibits a progressively more dispersed spatial distribution and a decreasing temperature throughout its evolution.

\begin{figure}[tp!]
\includegraphics[width=\linewidth]{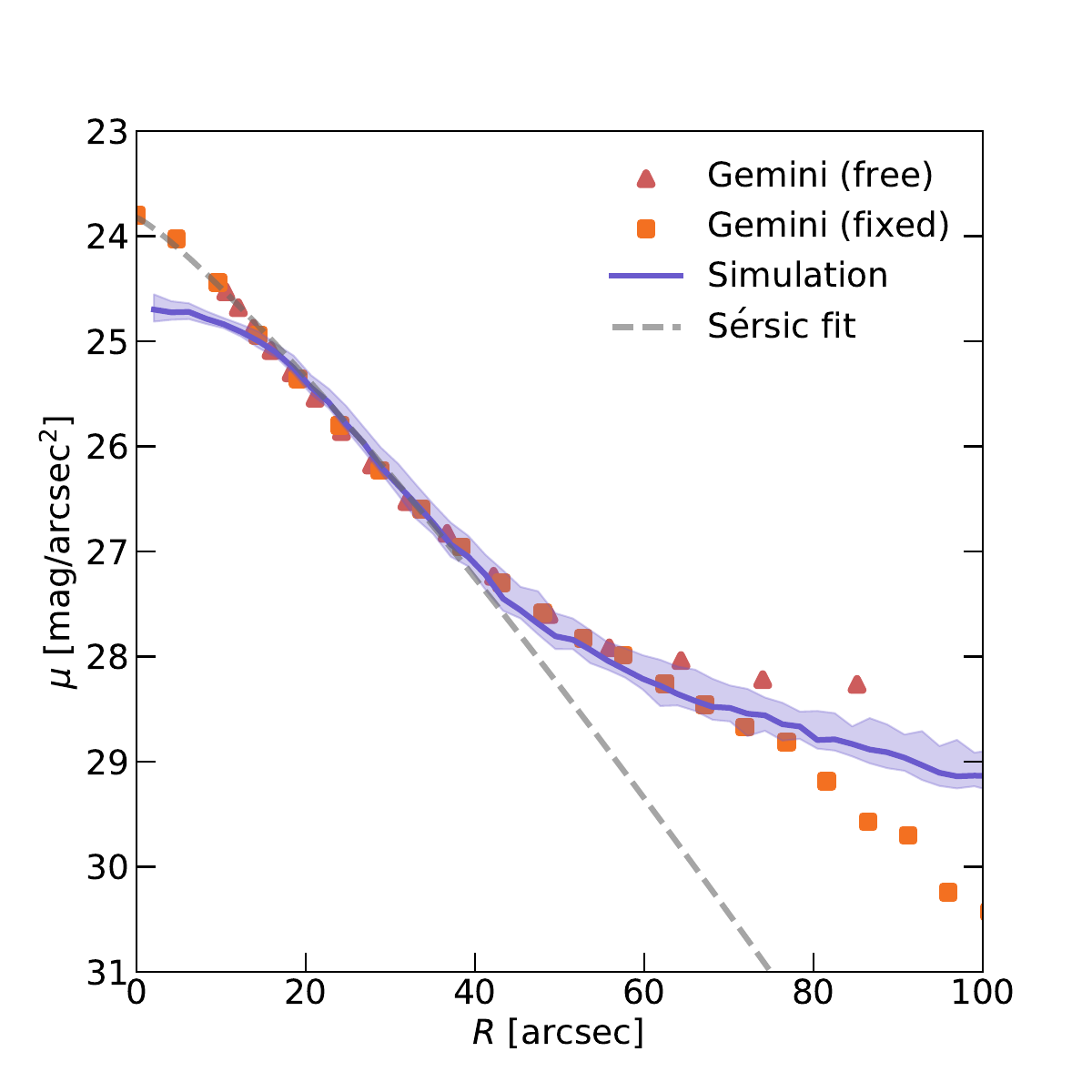}
\caption{\label{star_profile}
The radial surface brightness profile of DF4.
The solid purple line represents the profile from our fiducial simulation.
The light purple region marks the error range caused by projection orientations, and the lower and upper margins figure the 15 and 85 percentile of the values.
The orange squares and red triangles denote the profiles from Gemini images obtained by fitting isophotes with fixed and free ellipses, respectively \cite{golini2024ultra}.
The dashed gray line represents a S$\rm \acute{e}$rsic profile with an effective radius of $R_{\rm e}=19.8''$ and an index of $n=0.85$, fitted from the innermost region of DF4 \cite{keim2022tidal}.}
\end{figure}

\begin{figure}[htp!]
\includegraphics[width=\linewidth]{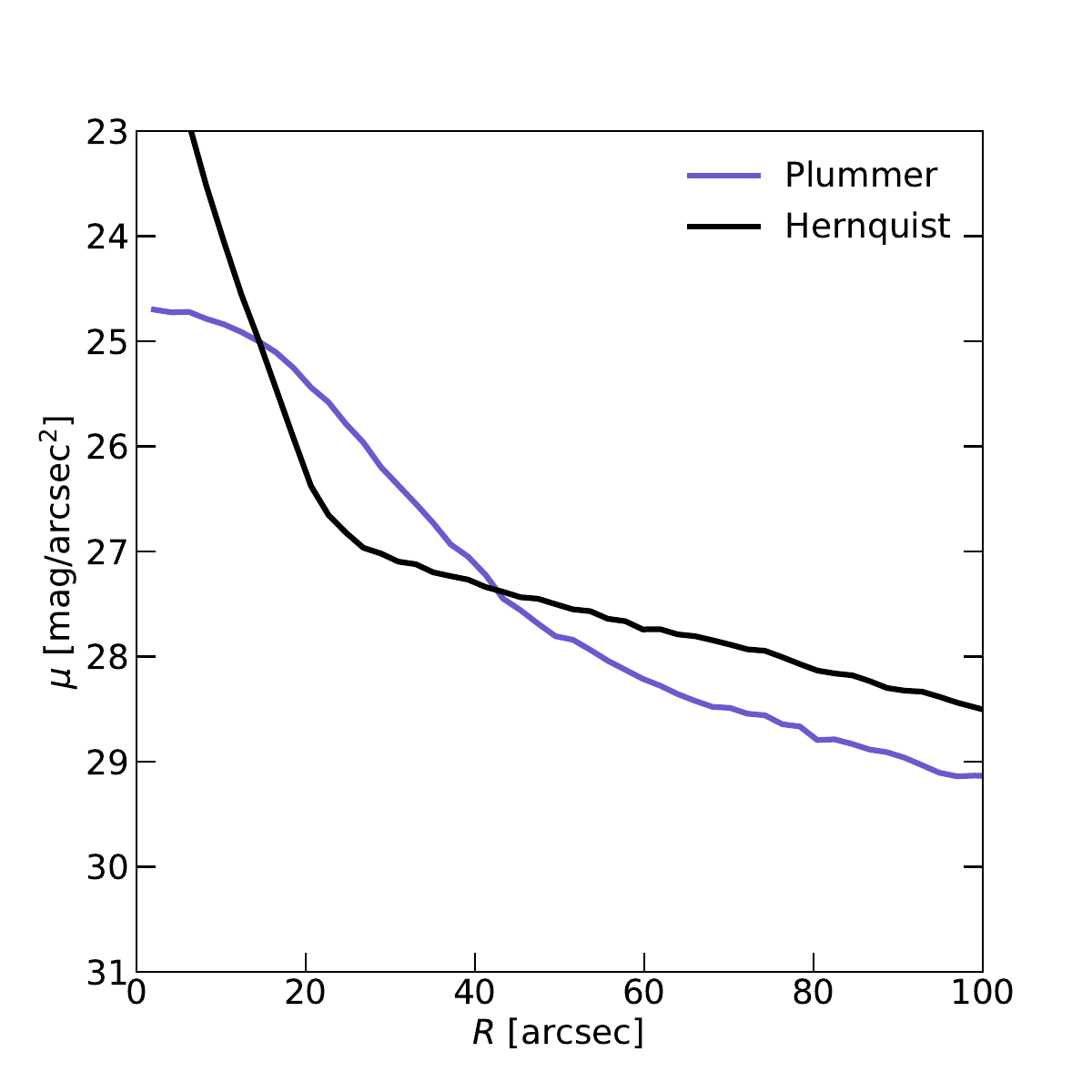}
\caption{\label{profile_contrast}
Comparison of radial surface brightness profiles obtained from different initial distributions.
Both profiles are generated at the time when results from the corresponding simulation best match all observational data, and are derived from density profiles using the same mass-to-light ratio.}
\end{figure}

\subsection{\label{subsec:surface_brightness}Surface brightness Profile}

Fig. \ref{star_profile} demonstrates the radial surface brightness profile of DF4 produced by our fiducial simulation at $t=-0.34~\rm Gyr$, along with the observational data from Gemini ultra-deep imaging.
Our profile assumes circular isophotes, with the shaded region representing results from various projection directions.
By selecting an appropriate mass-to-light ratio and setting the distance of DF4 as $D=20~\rm Mpc$, the density profile can be converted into a surface brightness profile.
Note that the value of the mass-to-light ratio does not change the shape of the profile, and only results in a vertical shift.
For comparison, we also show the two Gemini profiles provided by \cite{golini2024ultra}, which are obtained by fitting isophotes using fixed and free ellipses.
In the region with $R<60''$, where the ellipticity is small and the differences between these two profiles are minimal, the surface brightness profile from our fiducial simulation demonstrates a robust consistency with the Gemini results.

The distortion radius $R_{\rm distortion}$ marks the location where tidal distortion arises and tidal tails emerge. This radius can be identified by locating the position where the ellipticity of isophotes shows a sudden rise.
Using this method, Ref. ~\cite{keim2022tidal} and \cite{golini2024ultra} suggested a distortion radius of $R_{\rm distortion}=39''$ and $45''$, respectively.
The surface brightness profile produced by our simulation exhibits a change in slope at $R=43''$, indicating the appearance of tidal tails and aligning well with the distortion radii derived from observations.
Additionally, Ref.~\cite{Montes:2020zaz} and \cite{keim2022tidal} reported a break radius of $R_{\rm break}=43.8''$ and $46''$, respectively.
Beyond this break radius, the surface brightness exhibits an excess of at least $0.2~\rm mag/arcsec^{2}$ compared to the S$\rm \acute{e}$rsic profile fitted from the innermost region of DF4.
Our profile yields a break radius of $R_{\rm break}=45.4''$, consistent with the reported break radii.
Therefore, our fiducial simulation accurately reproduces the essential features of tidal tails in DF4.

The discrepancy in the central region with $R<12''$ between our simulated profile and the observational data can be attributed to the choice of the initial Plummer profile.
The 3D Plummer profile has a density core in the central region, where the impact of tidal effects is relatively minor.
Consequently, the final surface brightness profile retains a lower slope in this innermost region.
To illustrate the impact of the initial distribution on the final profile, we conduct a simulation using an Hernquist profile \cite{Hernquist:1990be} with an effective radius $R_{\rm e}=1.45~\rm kpc$ as the initial condition for the star population, while maintaining all other conditions identical to the fiducial simulation.
The resulting surface brightness profile of this new simulation at $t=-0.98~\rm Gyr$, when the simulation results best align with observational data, is shown in Fig. \ref{profile_contrast}.
The Hernquist profile exhibits a log-slope of $\gamma=-1$ in the central region, sharper than the Plummer profile, leading to a final profile with higher luminosity and a steeper slope in the innermost region.
The steeper density profile of the Hernquist distribution provides greater resistance to tidal forces, allowing for a larger effective radius to achieve the desired DM-to-star mass ratio consistent with observations.
In this study, we take the Plummer profile as the initial star distribution for simplicity.
A S$\rm \acute{e}$rsic profile with an appropriate index may yield a final surface brightness profile more in line with the observations, and mitigate the issue of using a small effective radius in our fiducial simulation.

\section{\label{sec:summary}Summary}
Our study demonstrates that a DF4-like galaxy can form through tidal stripping in the DM halo of NGC1052.
Our initial conditions of the satellite system are close to the tails of corresponding distributions, i.e. a diffuse DM halo and a compact star population.
This could be reasonable since a DM-deficient galaxy with obvious tidal tails is rare in the universe.
Even with this initial satellite system, a strong DM self-interaction is still required to attain the extremely low DM-to-star mass ratio inferred from observations.
Combined with two contrast simulations, our fiducial simulation illustrates the substantial influence of SIDM on reducing the mass ratio.
With tidal effects enhanced by SIDM, the tidal evolution produces a radial surface brightness profile which closely resembles that from deep images apart from the innermost region.
This profile accurately captures tidal features, including the diffuse effective radius and the presence of tidal tails.
Our work shows that the tidal formation of DF4 favors a strong DM self-interaction, which may provide insights into the nature of DM.
\\
\\

\acknowledgments
This work is supported by the National Natural Science Foundation of China under grant No. 12175248.

\bibliography{apssamp}

\end{document}